\documentstyle[12pt]{article}
\input amssym.def
\begin{document}
\def\be{\begin{equation}}
\def\ee{\end{equation}}
\def\ba{\begin{array}}
\def\ea{\end{array}}
\def\Cb{{\Bbb C}}
\def\ra{\rangle}
\def\la{\langle}

\centerline{\Large\bf A Complete Set of Local Invariants for a
Family of } \vspace{2ex} \centerline{\Large\bf Multipartite Mixed
States} \vspace{4ex}

\begin{center}
{\normalsize Xiao-Hong Wang$^1$,  Shao-Ming Fei$^{1,\  2}$  and Ke
Wu$^1$ }
\end{center}

\begin{center}
\begin{minipage}{5.3in}
{\small \sl $^1$ Department of Mathematics, Capital  Normal
University, Beijing, China}

{\small \sl $^2$ Max-Planck-Institute for Mathematics in the
Sciences, 04103 Leipzig, Germany}

\end{minipage}
\end{center}

\begin{center}
\begin{minipage}{5.2in}

 \vskip 1mm PACS number(s): 03.67.-a, 02.20.Hj, 03.65.-w
\end{minipage}
\end{center}

\begin{abstract}
We study the equivalence of quantum states under local unitary
transformations by using the singular value decomposition. A
complete set of invariants under local unitary transformations is
presented for several classes of tripartite mixed states in ${\Cb}^K
\otimes {\Cb}^M \otimes {\Cb}^N$ composite systems. Two density
matrices in the same class are equivalent under local unitary
transformations if and only if all these invariants have equal
values for these density matrices.
\end{abstract}

\section{Introduction}

Quantum entanglement is playing very important roles in quantum
information processing. Quantum entangled states are the key
resource in quantum information processing \cite{Nie00} such as
teleportation, super-dense coding, key distribution, error
correction and quantum repeater. Therefore it is of great importance to
classify and characterize the quantum states.

The nature of the entanglement among the parts of a composite system
does not depend on the labeling of the basis states of the
individual subsystems. It is therefore invariant under unitary
transformations of the individual state spaces. Such transformations
are referred to as local unitary transformations.
The polynomial invariants of local unitary transformations have been
discussed in \cite{VW,Sudbery,BL}.
General methods, which allow in principle to compute all such
invariants, but are in fact not really operational, were introduced in
\cite{p27,p19,p29,p5}.
More explicit complete and partial solutions have been found for some special cases:
two qubits \cite{makhlin} and three qubits \cite{Linden99,sun} systems,
three qutrits \cite{p10}, generic mixed states \cite{p4},
special families of tripartite pure qudits \cite{ACFW2005-1,ACFW2005-2,W2006}.

The problem of classifying states under local unitary
transformations can been solved completely for bipartite pure states. As the
set of Schmidt coefficients forms a complete set of invariants under
local unitary transformations, two bipartite pure states are
equivalent under local unitary transformations if and only if they
have the same Schmidt coefficients.
For multiple composite system, there does not exist Schmidt
decomposition in general. There are different generalizations for
Schmidt decomposition in multipartite quantum pure states \cite{CHS,
AACJLT,Peres,abls,Thapliyal}, but the results are not sufficient to solve the
local equivalence problem. For multipartite mixed states,
much less is known about the equivalence under local unitary transformations.

Another classification of quantum states is the one under
stochastic local operations and classical communications (SLOCC).
Invariants under SLOCC have been also extensively studied \cite{LT,LP,Miyake}.
Recently, Lamata et al. \cite{LLSS} used the method of
singular value decomposition and presented an inductive classification of
multipartite qubit systems under SLOCC.

In this letter, we study the equivalence of multipartite mixed states under
local unitary transformations by using the singular value
decomposition. Let ${\cal H}_1$ (resp. ${\cal H}_2$) be $M$ (resp. $N$) dimensional
complex Hilbert spaces ($M\leq N$).  A mixed state $\rho$ in ${\cal
H}_1\otimes {\cal H}_2$ with rank $r(\rho)=n\leq M^2$ can be
decomposed according to its eigenvalues $\lambda_i$ and eigenvectors
$\vert\nu_i\rangle$, $i=1,...,n$:
$$
\rho=\sum_{i=1}^n\lambda_i\vert\nu_i\rangle\langle\nu_i\vert.
$$

In \cite{WACF2006}, a class of bipartite mixed states $\Gamma_0$ has
been defined: $\Gamma_0$ contains all the states $\rho$ in ${\cal
H}_1\otimes {\cal H}_2$ satisfying \be\label{comm} [\rho_i,
\rho_j]=0, ~~[\theta_i, \theta_j]=0,~~~i,j=1,2,\cdots,n, \ee where
$\rho_i$ are full rank matrices,
$$
\rho_i=Tr_2 \vert\nu_i\rangle\langle\nu_i\vert,~~~
\theta_i=(Tr_1 \vert\nu_i\rangle\langle\nu_i\vert)^\ast, ~~~
i=1,...,n,
$$
$Tr_1$ (resp. $Tr_2$) denotes the partial trace over ${\cal H}_1$
(resp. ${\cal H}_2$). We denote by $^\dag$, $^*$
and $^t$ the adjoint, complex conjugation and transposition,
respectively.

It has been shown that two mixed states in $\Gamma_0$ are equivalent under local unitary
transformations if and only if the following invariants ((a) or
(b))have the same values for both mixed states \cite{WACF2006}:
$$\begin{array}{l}
(a)~~~~Tr(\rho_i^\alpha), ~~~Tr(\rho^\gamma),~~~
\alpha=1,2,\cdots,M,~~~\gamma=1,2,\cdots,MN.\\[3mm]
(b)~~~Tr(\theta_i^\beta), ~~~Tr(\rho^\gamma),~~~
\beta=1,2,\cdots,N,~~~\gamma=1,2,\cdots,MN.
\end{array}$$

The set of such states in $\Gamma_0$ is not trivial.
In fact, $\Gamma_0$ is a subset of  the Schmidt-correlated (SC)
states \cite{Virmani_sacchi}. The Schmidt-correlated (SC) states are
defined as mixtures of pure states, sharing the same Schmidt bases.
It was first appeared in \cite{Rains}, named as maximally correlated
state. For Schmidt-correlated states, for any classical measurement,
two observers Alice and Bob will always obtain the same result. Two
SC states can always be optimally discriminated locally. It is
interesting that maximally entangled states (Bell state) can always
be expressed in Schmidt correlated form.  SC states naturally appear
in a bipartite system dynamics with additive integrals of motion
\cite{Khasin06}. Hence, these states form an important class of
mixed states from a quantum dynamical perspective. From the
definition of SC state, we know the states in $\Gamma_0$ are all SC
states. Therefore we can judge whether a state in $\Gamma_0$ is
separable or not by calculating the negativity of this
state \cite{KKS}.

Here we give a simple way to construct some families of states
in $\Gamma_0$.  For $M=N=4$, one can set
$|\psi_1\ra=(|00\ra+|12\ra+|21\ra+|33\ra)/2$ and
$|\psi_2\ra=(|01\ra+|10\ra+|23\ra+|32\ra)/2$, where $|ij\ra$,
$i=0,1,...,M-1$, $j=0,1,...,N-1$, are the basis of ${\cal
H}_1\otimes {\cal H}_2$. Then $\rho=\alpha |\psi_1\ra\la\psi_1|
+(1-\alpha)|\psi_2\ra\la\psi_2|$ is a rank two state belonging to
$\Gamma_0$ for $0<\alpha< 1$. For general even $M=N=d+1$, a state
$\rho=\alpha |\psi_1\ra\la\psi_1| +(1-\alpha)|\psi_2\ra\la\psi_2|$
 is in $\Gamma_0$, where
$|\psi_1\ra=(|00\ra+|12\ra+|21\ra+|34\ra+|43\ra+...+|dd\ra)/\sqrt{M}$
and
$|\psi_2\ra=(|01\ra+|10\ra+|23\ra+|32\ra+...+|d-1,d\ra+|d,d-1\ra)/\sqrt{M}$.

For $M=N=5$, one can set
$|\phi_1\ra=(|00\ra+|12\ra+|21\ra+|34\ra+|43\ra)/\sqrt{5}$ and
$|\phi_2\ra=(|01\ra+|10\ra+|23\ra+|32\ra+|44\ra)/\sqrt{5}$. Then
$\rho=\alpha |\phi_1\ra\la\phi_1| +(1-\alpha)|\phi_2\ra\la\phi_2|$
is a rank two state in $\Gamma_0$. For general odd $M=N$,
$|\phi_1\ra$ and $|\phi_2\ra$ can be similarly constructed.

We can also construct higher rank states in $\Gamma_0$. For example,
for $M=N=4$, by adding $|\psi_3\ra=(|11\ra+|02\ra+|20\ra+|33\ra)/2$,
we have that
$\rho=\alpha|\psi_1\ra\la\psi_1|+\beta|\psi_2\ra\la\psi_2|+(1-\alpha-\beta)
|\psi_3\ra\la\psi_3|$ is a state in $\Gamma_0$. For odd $M=N=5$, we
have $|\phi_3\ra=(|04\ra+|13\ra+|22\ra+|31\ra+|40\ra)/\sqrt{5}$ and
$\rho=\alpha|\phi_1\ra\la\phi_1|+\beta|\phi_2\ra\la\phi_2|+(1-\alpha-\beta)
|\phi_3\ra\la\phi_3|~\in \Gamma_0$.

The states constructed above are all distillable. The rank of
reduced density matrices, which are in fact identity matrices, are
greater than the rank of $\rho$ itself. They are all NPPT (non
positive partial transpose) entangled states.

\section{ Tripartite Quantum Pure States}

We first discuss the locally invariant properties of arbitrary
dimensional tripartite pure quantum states. Let ${\cal H}_1$, ${\cal
H}_2$ and ${\cal H}_3$ be  $K$, $M$ and $N$ dimensional complex
Hilbert spaces with the orthonormal bases $\{\vert
e_i\rangle\}_{i=1}^{K}$ , $\{\vert f_i\rangle\}_{i=1}^{M}$ and
$\{\vert h_i\rangle\}_{i=1}^{N}$, respectively.

$|\Psi\ra$ can be regarded as a bipartite state by taking ${\cal
H}_1$ (resp. ${\cal H}_2$, ${\cal H}_3$) and ${\cal H}_2\otimes
{\cal H}_3$ (resp. ${\cal H}_1\otimes {\cal H}_3$, ${\cal
H}_1\otimes {\cal H}_2$) as the two subsystems. We denote these
three bipartite decompositions as $1-23$ (resp. $2-13$, $3-12$).
Let $a_{ijk}$  be the coefficients of $|\Psi\ra$ in orthonormal
bases $\vert e_i\rangle\otimes \vert f_j\rangle\otimes \vert
h_k\rangle$. Let $A_1$ (resp. $A_2$, $A_3$) denote the matrix with respect to the
bipartite state in $1-23$ (resp. $2-13$, $3-12$) decomposition, i.e.
taking the subindices $i$ (resp. $j$, $k$) and $jk$ (resp. $ik$,
$ij$) of $a_{ijk}$ as the row and column indices of $A_1$ (resp. $A_2$,
$A_3$).

Taking partial trace of $|\Psi\ra\la\Psi|$ over the respective
subsystems, we have $\tau_1=Tr_1|\Psi\ra\la\Psi|=A^t_1A^*_1$,
 $\tau_2=Tr_2|\Psi\ra\la\Psi|=A^t_2A^*_2$,
 $\tau_3=Tr_3|\Psi\ra\la\Psi|=A^t_3A^*_3$.
The reduced matrices $\tau_1$, $\tau_2$ and $\tau_3$ can be decomposed according to their
 eigenvalues and eigenvectors, e.g.
$$
\tau_1=\sum_{i=1}^{n_1}\lambda_i^1\vert\nu_i^1\rangle\langle\nu_i^1\vert,
$$
where $\lambda_i^1$ resp. $\vert\nu_i^1\rangle$, $i=1,...,n_1$, are
the nonzero eigenvalues resp. eigenvectors of the density matrix
$\tau_1$.

Let $A_i^1$ denote the matrix with entries given by the coefficients of
$\vert\nu_i^1\rangle$ in the bases $\vert
f_k\rangle\otimes \vert h_l\rangle$. We have
$$\rho_i^1=Tr_3
\vert\nu_i^1\rangle\langle\nu_i^1\vert={A_i^1}{A_i^1}^\dag,~~~
\theta_i^1=(Tr_2
\vert\nu_i^1\rangle\langle\nu_i^1\vert)^\ast={A_i^1}^\dag {A_i^1},
~~~ i=1,...,n_1. $$

Set
$$ I_{\alpha}^1(|\Psi\ra)
=Tr(\rho_i^1)^{\alpha},~~~\alpha=1,2,\cdots,M, $$
$$\label{invariant1} J_{\beta}^1(|\Psi\ra)
=Tr(\theta_i^1)^{\beta},~~~\beta=1,2,\cdots,N,
$$
$$K_{\gamma}^1(|\Psi\ra)=Tr(\tau_1^\gamma),~~~\gamma=1,2,\cdots,MN.$$

It is easy to prove that $I_{\alpha}^1(|\Psi\ra)$,
$J_{\beta}^1(|\Psi\ra)$ and $K_{\gamma}^1(|\Psi\ra)$
 are all invariants
under local unitary transformations.

Let $\Gamma_1$ denote a class of tripartite pure states $|\Psi\ra$
satisfying \be\label{commpp} [\rho_i^1, \rho_j^1]=0, ~~[\theta_i^1,
\theta_j^1]=0 \ee with $\rho_i^1$ being full rank matrices,
$i,j=1,2,\cdots, n_1$.

\noindent{\sf [Theorem 1]} Two pure states in $\Gamma_1$
are equivalent under local unitary transformations if and only if
the following invariants ((c) or (d)) have the same values for both states:
$$\ba{l}(c)~~~I_{\alpha}^1(|\Psi\ra),~~~
K_{\gamma}^1(|\Psi\ra),~~~\alpha=1,2,\cdots,M,~~~\gamma=1,2,\cdots,MN.
 \\[3mm]
(d)~~~J_{\beta}^1(|\Psi\ra),~~~K_{\gamma}^1(|\Psi\ra),
~~~\beta=1,2,\cdots,N, ~~~\gamma=1,2,\cdots,MN. \ea$$

We only need to prove the sufficient part. Assume
$|\Psi\ra,~|\Psi^\prime\ra\in\Gamma_1$.
$K_{\gamma}^1(|\Psi\ra)=K_{\gamma}^1(|\Psi^\prime\ra)$ imply that
$A_1$ and $A_1^\prime$ have the same singular values, therefore
there exists unitary matrices $U_1$ and $U_{23}$ such that
$|\Psi'\rangle=U_1\otimes U_{23}|\Psi\rangle$. If
$I_{\alpha}^1(|\Psi\ra)=I_{\alpha}^1(|\Psi^\prime\ra)$ or
$J_{\beta}^1(|\Psi\ra)=J_{\beta}^1(|\Psi^\prime\ra)$ holds, then
$\tau_1$ and $\tau_1^\prime$ are equivalent under local unitary
transformations by the sufficient condition of equivalence for
bipartite states under local unitary transformations. While in
\cite{ACFW2005-2} it has been proven that if
$|\Psi'\rangle=U_1\otimes U_{23}|\Psi\rangle$, with
$U_1\in{\mathrm{U(}}{\mathcal{H}}_1{\mathrm{)}}$,
$U_{23}\in\mathrm{U(}{\mathcal{H}}_2\otimes{\mathcal{H}}_3\mathrm{)}$
and ${Tr}_1\left(|\Psi'\rangle\langle\Psi'|\right)=
U_2\otimes U_3 {Tr}_1\left(|\Psi\rangle\langle\Psi|\right) U_{2}^{\dagger}\otimes
U_{3}^{\dagger}$, where
$U_2\in{\mathrm{U(}}{\mathcal{H}}_2\mathrm{)}$ and
$U_3\in{\mathrm{U(}}{\mathcal{H}}_3\mathrm{)}$,
then there exist matrices $V_1\in{\mathrm{U(}}{\mathcal{H}}_1\mathrm{)}$,
$V_2\in{\mathrm{U(}}{\mathcal{H}}_2\mathrm{)}$,
$V_3\in{\mathrm{U(}\mathcal{H}}_3\mathrm{)}$ such that
$|\Psi'\rangle=V_1\otimes V_2\otimes V_3|\Psi\rangle$, i.e.,
$|\Psi\rangle$ and $|\Psi'\rangle$ are equivalent under local
unitary transformations. \hfill $\Box$

Let us consider for example two states
$|\Psi\ra=\sqrt{\frac{p}{3}}(|000\ra+|012\ra+|021\ra)+
\sqrt{\frac{1-p}{3}}(|101\ra+|110\ra+|122\ra)$ and
$|\Psi^\prime\ra=\sqrt{\frac{p}{3}}(|000\ra+|011\ra+|022\ra)+
\sqrt{\frac{1-p}{3}}(|101\ra+|112\ra+|120\ra)$ in ${\cal H}_1\otimes
{\cal H}_2\otimes {\cal H}_3, $ for the case $K=2$, $M=N=3$. It is
direct to verify that they are all states in $\Gamma_1$ with
$\rho_i^1=\theta_i^1=\frac{1}{3}I$, $i=1,2.$ As $\tau_1$ and
$\tau_1^\prime$ have the same eigenvalues, relation
$K_{\gamma}^1(|\Psi\ra)=K_{\gamma}^1(|\Psi^\prime\ra)$ holds, from
which and the following equations
$$Tr(\rho_i^1)=Tr({\rho_i^\prime}^1)=1,~~~
Tr({\rho_i^1})^2=Tr({{\rho_i^\prime}^1})^2=\frac{1}{3},
$$
by Theorem 1 we have that $|\Psi\ra$ and $|\Psi^\prime\ra$
are equivalent under local unitary transformations.  The same
results can be also obtained from
$K_{\gamma}^1(|\Psi\ra)=K_{\gamma}^1(|\Psi^\prime\ra)$ and the
following facts:
$$Tr(\theta_i^1)=Tr({\theta_i^\prime}^1)=1,~~~
 ~~~Tr({\theta_i^1})^2=Tr({{\theta_i^\prime}^1})^2=\frac{1}{3}.$$

As an alternative example we consider two states
$|\Psi\ra=\sqrt{\frac{\alpha}{3}}(|000\ra+|012\ra+|021\ra)+
\sqrt{\frac{\beta}{3}}(|101\ra+|110\ra+|122\ra)
+\sqrt{\frac{\gamma}{3}}(|202\ra+|211\ra+|220\ra)$ and
$|\Psi^\prime\ra=\sqrt{\frac{\alpha}{3}}(|000\ra+|011\ra+|022\ra)+
\sqrt{\frac{\beta}{3}}(|101\ra+|112\ra+|120\ra)+
\sqrt{\frac{\gamma}{3}}(|202\ra+|210\ra+|221\ra)$ in ${\cal
H}_1\otimes {\cal H}_2\otimes {\cal H}_3, $ with $K=M=N=3$,
$\alpha,\beta,\gamma\in R$, $\alpha+\beta+\gamma=1$. One can prove
that they are all states in $\Gamma_1$ with
$\rho_i^1=\theta_i^1=\frac{1}{3}I$, $i=1,2,3$, and  $\tau_1$,
$\tau_1^\prime$ have the same eigenvalues. As
$$Tr(\rho_i^1)=Tr({\rho_i^\prime}^1)=1,~~~
Tr({\rho_i^1})^2=Tr({{\rho_i^\prime}^1})^2=\frac{1}{3},~~~
Tr({\rho_i^1})^3=Tr({{\rho_i^\prime}^1})^3=\frac{1}{9},
$$
from Theorem 1 we have $|\Psi\ra$ and $|\Psi^\prime\ra$ are
equivalent under local unitary transformations. Moreover by using
the generalized concurrence \cite{FGWWW}, we have $C^3_3\neq 0,$
hence $|\Psi\ra$ and $|\Psi^\prime\ra$ are entangled.

{\sf [Remark]} We can also similarly define the set of states
$\Gamma_2$. Let $\tau_2$ be a reduced density matrix by tracing
$|\Psi\ra\la\Psi|$ over the second system. $\tau_2$ can be
decomposed according to its eigenvalues and eigenvectors:
$$
\tau_2=\sum_{i=1}^{n_2}\lambda_i^2\vert\nu_i^2\rangle\langle\nu_i^2\vert,
$$
where $\lambda_i^2$ resp. $\vert\nu_i^2\rangle$, $i=1,...,n_2$, are
the nonzero eigenvalues resp. eigenvectors of the density matrix
$\tau_2$. Define $ \left\{\rho_{i}^2\right\}$,
$\left\{\theta_{i}^2\right\}$,
$$\rho_i^2=Tr_3
\vert\nu_i^2\rangle\langle\nu_i^2\vert,~~~ \theta_i^2=(Tr_1
\vert\nu_i^2\rangle\langle\nu_i^2\vert)^\ast, ~~~ i=1,...,n_2. $$ We
define $\Gamma_2$ to be a set of tripartite pure states $|\Psi\ra$
satisfying \be\label{commB} [\rho_i^2, \rho_j^2]=0, ~~[\theta_i^2,
\theta_j^2]=0 \ee with $\rho_i^2$ being full rank matrices. Then we
also have the similar result:

\noindent{\sf [Theorem 2]} Two pure states in $\Gamma_2$
are equivalent under local unitary transformations if and only if
the following invariants ((e) or (f)) have the same values for both states:
$$\ba{l}(e)~~~I_{\alpha}^2(|\Psi\ra),~~~
K_{\gamma}^2(|\Psi\ra),~~~\alpha=1,2,\cdots,K,~~~\gamma=1,2,\cdots,KN,
 \\[3mm]
(f)~~~J_{\beta}^2(|\Psi\ra),~~~K_{\gamma}^2(|\Psi\ra),
~~~\beta=1,2,\cdots,N, ~~~\gamma=1,2,\cdots,KN, \ea$$ where
$I_{\alpha}^2(|\Psi\ra)=Tr(\rho_i^2)^{\alpha}$,
$J_{\beta}^2(|\Psi\ra) =Tr(\theta_i^2)^{\beta},$
$K_{\gamma}^2(|\Psi\ra)=Tr(\tau_2^\gamma).$

The set of states $\Gamma_3$ can be defined in a similar way and the
corresponding theorem (like theorem 1 and 2) can be obtained
similarly.

The results above can be generalized to general many partite
systems. As each $n$ partite pure states can be treated as a
bipartite one: the $j$th system and rest $n-1$ partite system, by
using the results of Lemma 2 in \cite{ACFW2005-2}, one can similarly
obtain a complete set of invariants for some classes of multipartite
pure states.

\section{ Tripartite Quantum Mixed States}

We consider now mixed states in ${\cal H}_1\otimes {\cal H}_2\otimes
{\cal H}_3$. We assume $K\le M,N$. Let $\rho$ be a density matrix
defined on ${\cal H}_1\otimes {\cal H}_2\otimes {\cal H}_3$ with
$r(\rho)=n\leq K^3$. $\rho$ can be decomposed according to its
eigenvalues and eigenvectors:
$$
\rho=\sum_{i=1}^n\lambda_i\vert\nu_i\rangle\langle\nu_i\vert,
$$
where $\lambda_i$ resp. $\vert\nu_i\rangle$, $i=1,...,n$, are the
nonzero eigenvalues resp. eigenvectors of the density matrix $\rho$.
We introduce
\[ \rho_{i}=Tr_1 \vert\nu_i\rangle\langle\nu_i\vert,~~~
\theta_{i}=Tr_2 \vert\nu_i\rangle\langle\nu_i\vert,~~~
\gamma_{i}=Tr_3 \vert\nu_i\rangle\langle\nu_i\vert.\]

If we treat $\vert\nu_i\rangle$ as a bipartite state
$\vert\omega_i\rangle$ in $1-23$ system, let $A_{1i}$ denote the
matrix with entries given by the coefficients of
$\vert\omega_i\rangle$ in the bases $\vert e_k\rangle\otimes \vert
g_l\rangle,$ where $\vert g_{l}\rangle=\vert f_t\rangle \otimes
\vert h_s\rangle$, $l=ts$; $t=1,\cdots, M$, $s=1,\cdots, N.$
According to the result of bipartite system, we have
$$Tr_2
\vert\omega_i\rangle\langle\omega_i\vert=A_{1i}A_{1i}^\dag,~~~ (Tr_1
\vert\omega_i\rangle\langle\omega_i\vert)^\ast=A_{1i}^\dag A_{1i},
~~~ i=1,...,n. $$

As $\label{1f}Tr_2\vert\omega_i\rangle\langle\omega_i\vert
=Tr_3(Tr_2 \vert\nu_i\rangle\langle\nu_i\vert)$ and
$\label{2-f}Tr_1\vert\omega_i\rangle\langle\omega_i\vert =Tr_1
\vert\nu_i\rangle\langle\nu_i\vert$, we have
$$
\theta_{i}^{23}=A_{1i}A_{1i}^{\dag},~~~
\rho_{i}=(A_{1i}^{\dag}A_{1i})^*,
$$
where $  \theta_{i}^{23}=Tr_3(Tr_2
\vert\nu_i\rangle\langle\nu_i\vert).$

$\rho_i$ can be again decomposed according to its eigenvalues and
eigenvectors:
$$
\rho_i=\sum_{j=1}^{m_i}\alpha_j^i\vert\mu^i_j\rangle\langle\mu^i_j\vert,
$$
where $\alpha^i_j$ resp. $\vert\mu^i_j\rangle$, $j=1,...,m_i$, are
the nonzero eigenvalues resp. eigenvectors of the reduced density
matrix $\rho_i$.
Let $B^i_j$ denote the matrix with entries given by
coefficients of $\vert\mu^i_j\rangle$ in the bases
$\vert f_k\rangle\otimes \vert h_l\rangle$.
We further introduce $ \left\{\xi^i_j\right\}$, $
\left\{\eta^i_j\right\}$,
$$\xi^i_j=Tr_3
\vert\mu^i_j\rangle\langle\mu^i_j\vert=B^i_j{B^i_j}^\dag,~~~
\eta^i_j=(Tr_2
\vert\mu^i_j\rangle\langle\mu^i_j\vert)^\ast={B^i_j}^\dag B^i_j, ~~~
j=1,...,m_i. $$

Let $\Gamma$ denote a class of tripartite mixed states satisfying
\be\label{comm-1} [\rho_i, \rho_k]=0, ~~ ~~[\theta_i^{23},
\theta_k^{23}]=0 \ee with $\theta_i^{23}$ being full rank matrices,
$i,k=1,2,\cdots, n$, and \be\label{comm-2} [\xi^i_t, \xi^k_l]=0, ~~
~~[\eta^i_t, \eta^k_l]=0 \ee with $\xi^i_t$ being full rank
matrices, $\forall i,k=1,2,\cdots,n$, $t=1,2,\cdots,m_i$,
$l=1,2,\cdots,m_k.$

\noindent{\sf [Theorem 3]} Two mixed states in $\Gamma$ are equivalent
under local unitary transformations if and only if the following
invariants ((g) or (h)) have the same values for both mixed states:
$$\begin{array}{l}
(g)~~~Tr(\rho_i)^{\alpha},~~~ Tr(\xi^k_l)^{\alpha}, ~~~
Tr(\rho^\gamma), ~~~\alpha=1,2,\cdots,M,~~~\gamma=1,2,\cdots,MN.
\\[3mm]
(h)~~~Tr(\theta_i^{23})^{\beta}, ~~~Tr(\eta^k_l)^{\beta},
~~~Tr(\rho^\gamma), ~~~\beta=1,2,\cdots,N,~~~\gamma=1,2,\cdots,MN.
\end{array}$$

\noindent{\sf [Proof]:} If $\rho$ and $\rho^\prime\in\Gamma$
are equivalent under the local unitary
transformation $u\otimes v\otimes w$, $\rho^\prime=u\otimes v\otimes
w ~\rho~u^\dag\otimes v^\dag\otimes w^\dag$, then
$\vert\nu_i^\prime\rangle=u\otimes V\vert\nu_i\rangle$, where
$V=v\otimes w$, namely $A_{1i}$ is mapped to $A_{1i}^\prime=u A_{1i}
V^t$. Therefore
$$\label{aa}
{\theta^\prime}_i^{23}=A_{1i}^\prime A_{1i}^{\prime\dag}=u A_{1i}
A_{1i}^\dag u^\dag=u\theta_i^{23} u^\dag,$$
$$\rho_i^\prime=(A_{1i}^{\prime\dag}A_{1i}^\prime)^\ast=V
(A_{1i}^\dag A_{1i})^\ast V^\dag =V \rho_i V^\dag=v\otimes w \rho_i
v^\dag\otimes w^\dag.
$$
Thus $\rho_i$ and $\rho_i^\prime$ are equivalent under the local
unitary transformation $v\otimes w$, from the results of bipartite
system \cite{WACF2006} we have $
Tr(\xi^k_l)^{\alpha}=Tr({\xi^\prime}^k_l)^{\alpha}$ and
$Tr(\eta^k_l)^{\beta}=Tr({\eta^\prime}^k_l)^{\beta}$. Therefore
$(g)$ and $(h)$ hold.

Conversely, $Tr(\rho^\gamma)=Tr({\rho^\prime}^\gamma)$ imply that
$\rho$ and $\rho^\prime$  have the same eigenvalues. We now prove
that there exist common unitary matrices $V_1, V_2, V_3$ such that
$\vert\nu_i^\prime\rangle=V_1\otimes V_2\otimes
V_3\vert\nu_i\rangle\,$ by using Lemma 2 in \cite{ACFW2005-2}.

From the relation $Tr(\rho_i)^{\alpha}=Tr(\rho_i^\prime)^{\alpha}$
in (g) and the condition $(\ref{comm-1})$, we have common unitary
matrices $U_1$ and $U_{23}$ for all $i$ such that
$\vert\nu_i^\prime\rangle=U_1\otimes U_{23}\vert\nu_i\rangle$.

The relation $Tr(\xi^k_l)^{\alpha}=Tr({\xi^\prime}^k_l)^{\alpha}$ in
(g) and the condition $(\ref{comm-2})$ imply that $\rho_i$ and
${\rho_i^\prime}$ are equivalent under local unitary
transformations, $\rho_i^\prime=U_i\otimes V_i\rho_i U_i^\dag
\otimes V_i^\dag$ according to the results of bipartite system
\cite{WACF2006}. For the case $i\neq k$ in condition $(\ref{comm-2})$,
$(\ref{comm-2})$ implies that there exist common unitary matrices
$U$ and $V$ such that $\rho_i^\prime=U\otimes V\rho_i U^\dag \otimes
V^\dag$. To elucidate this we just show the case $n=2$. For a
rank-two state $\rho$ we have
$$
\rho_1=\sum_{j=1}^{m_1}\alpha_j^1\vert\mu^1_j\rangle\langle\mu^1_j\vert,
~~~~~
\rho_2=\sum_{j=1}^{m_2}\alpha_j^2\vert\mu^2_j\rangle\langle\mu^2_j\vert.
$$

$Tr(\xi^1_j)^{\alpha}=Tr({\xi^\prime}^1_j)^{\alpha}$ implies that
$\xi^1_j$ and ${\xi^\prime}^1_j$ are equivalent under unitary
transformations. Therefore $B^1_j$ and ${B^\prime}^1_j$ have the
same singular values. \be\label{co-1}[\xi^1_t, \xi^1_l]=0~~~\ee and
\be\label{co-2}[\eta^1_t, \eta^1_l]=0~~~\ee imply that ( from
singular value decomposition) there exist common unitary matrices
$U_1, U_1^\prime$ and $V_1, V_1^\prime$ such that \be
\label{co-3}U_1{B^1_j} V_1=U_1^\prime {B^\prime}^1_j V_1^\prime. \ee

While \be\label{co-4}[\xi^2_t, \xi^2_l]=0~~~\ee and
\be\label{co-5}[\eta^2_t, \eta^2_l]=0~~~\ee imply that there exist
common unitary matrices $U_2, U_2^\prime$ and $V_2, V_2^\prime$ such
that \be\label{co-6}U_2{B^2_j} V_2=U_2^\prime {B^\prime}^2_j
V_2^\prime.\ee

From $(\ref{co-1})$ and $(\ref{co-4})$, we have $U_1=U_2$. From
$(\ref{co-2})$ and $(\ref{co-5})$, we have $V_1=V_2$. Hence $
{B^\prime}^i_j =U {B^i_j} V^t$, and
$\vert{\mu^\prime}^i_j\rangle=U\otimes V \vert\mu^i_j\rangle$,
$j=1,...,m_i$. Therefore, $\rho_i^\prime=U\otimes V
~\rho_i~U^\dag\otimes V^\dag$. Hence $\rho_i^\prime$ and $\rho_i$
are equivalent under local unitary transformations.

Therefore, from Lemma 2 in \cite{ACFW2005-2} we have that tripartite
states $\vert\nu_i\rangle$ and $\vert\nu_i^\prime\rangle$ are
equivalent under local unitary transformations. In fact there exist
common unitary matrices $V_i, i=1,2,3$, such that
$\vert\nu_i^\prime\rangle=V_1\otimes V_2\otimes
V_3\vert\nu_i\rangle\,,$ where $V_1=WU_1$, $V_2=U$, $V_3=V$
($[\theta_i^{23}, \theta_j^{23}]=0$ imply that there exists common
$W$ for different $\nu_i$). Therefore, we have
$\rho^\prime=V_1\otimes V_2\otimes V_3\rho V_1^\dag \otimes V_2^\dag
\otimes V_3^\dag.$

Thus from (g) we get that $\rho$ and $\rho^\prime$ are equivalent
under local unitary transformations. One can similarly prove $\rho$
and $\rho^\prime$ are equivalent under local unitary transformations
from (h). $\hfill$$\Box$

We have discussed the local invariants for arbitrary dimensional
tripartite quantum mixed states in ${\Cb}^K \otimes {\Cb}^M \otimes
{\Cb}^N$ composite systems and have presented sets of invariants
under local unitary transformations for some classes of
tripartite mixed states. The invariants in a set is not
necessarily independent, but they are sufficient to judge if two
states in $\Gamma$ or $\Gamma_i$, $i=1,2,3$, are equivalent under
local unitary transformations. For three qubits case, $K=M=N=2$, a
set of invariants has been presented in \cite{Linden99,sun} for a
special class of states. By using  the method in
\cite{ACFW2005-1,ACFW2005-2}, the results can be generalized to
detect local equivalence for some special classes of general
multipartite states.

\vspace{1.0truecm}

\noindent {\bf Acknowledgments} The work is supported by Beijing
Municipal Education Commission (No. KM 200510028021, KM200510028022)
and NKBRPC(2004CB318000), NSFC project 10675086. X.H. Wang
gratefully acknowledges S. Albeverio for continuous
encouragement and L. Cattaneo for valuable discussion.

\end{document}